\documentclass[11pt,preprint]{aastex}

\newcommand\simlt{\lower.5ex\hbox{$\; \buildrel < \over \sim \;$}}
\newcommand\simgt{\lower.5ex\hbox{$\; \buildrel > \over \sim \;$}}

\begin{document}

\title{Structure and Nuclear Composition of General Relativistic, Magnetohydrodynamic Outflows from Neutrino-Cooled Disks}
\author{Yudith Barzilay and Amir Levinson\altaffilmark{1}}
\altaffiltext{1}{School of Physics \& Astronomy, Tel Aviv
University, Tel Aviv 69978, Israel; Levinson@wise.tau.ac.il}
\begin{abstract}
We compute the structure and degree of neutronization of general
relativistic magnetohydrodynamic (GRMHD) outflows originating from
the inner region of neutrino-cooled disks.  We consider both,
outflows expelled from a hydrostatic disk corona and outflows driven
by disk turbulence.  We show that in outflows driven thermally from
a static disk the electron fraction quickly evolves to its
equilibrium value which is dominated by neutrino capture.  Those
outflows are generally proton rich and, under certain conditions,
can be magnetically dominated.  They may also provide sites for
effective production of $^{56}$Ni. Centrifugally driven outflows and
outflows driven by disk turbulence, on the other hand, can preserve
the large in-disk neutron excess.  Those outflows are, quite
generally, subrelativistic by virtue of the large mass flux driven
by the additional forces.

\end{abstract}

\section{Introduction}

The structure and nuclear composition of outflows from a disk
accreting rapidly onto a central black hole is a topic of
considerable interest.  Such systems are believed to form following
the collapse of rotating massive stars in failed supernovae events
(collapsars), or the merger of neutron stars and black holes, and
there is some evidence that link them to GRBs.  For the range of
accretion rates ($\dot{M}_{acc}\sim0.01 - 10$ $M_\odot$ yr$^{-1}$)
and viscosity parameters ($\alpha_{\rm vis}\sim0.01 -0.1$)
anticipated, the disk surrounding the black hole is sustained at MeV
temperatures and its midplane density is high, in excess of
$10^{10}$ gr cm$^{-3}$ (e.g., Popham et al., 1999). Under such
conditions the inner parts of the disk, within about $10$ to $100
r_g$ depending on parameters, consist of neutron rich matter that
cools predominantly via emission of MeV neutrinos (Popham et al.
1999; Di Matteo et al. 2002; Kohri \& Mineshing 2002; Pruet at al.
2003; Chen \& Beloborodov 2007, hereafter CB07). For
$\dot{M}_{acc}\simgt 0.1$ and $\alpha_{\rm vis}\simlt 0.03$ the
neutron-to-proton ratio within $10 r_g$ exceeds $30$ (CB07).

Viscous and neutrino heating of the upper disk layers drives a powerful wind from the disk.
This wind is most likely baryon rich and expands at sub or mildly relativistic speeds
except, perhaps, inside a core containing the putative baryon poor GRB jet.
The baryon rich wind emanating from the disk is a potential site for efficient
nucleosynthesis (Pruet et al. 2003, 2004), and may account for the SN-like features observed in
several long GRBs.  It may also play an important role in the collimation of the central
GRB-producing jet (Levinson \& Eichler 2000; Alloy et al. 2000; Rosswog \& Ramirez-Ruiz 2003;
Alloy et al. 2005; Bromberg \& Levinson 2007).  The latter may be launched magnetically from
the very inner regions of the disk (e.g., Vlahakis \& Konigl 2003; Levinson 2006) or may be produced
via a Blandford-Znajek mechanism (Levinson \& Eichler 1993; Meszaros \& Rees 1997;
Van Putten 2001; Van Putten \& Levinson 2003).  If, as often assumed, the GRB jet picks up nuclei from the
disk (e.g., Derishev et al. 1999; Beloborodov 2003) then it may
contain matter with a large neutron excess that is likely to affect its dynamics
(Fuller et al. 2000; Vlahakis, Peng \& Konigl 2003; Rossi et al. 2005), as well as some of the
characteristics of the afterglow emission (e.g., Derishev et al. 1999; Bachall \& Meszaros 2000).

Whether disk outflows can preserve the large in-disk neutron excess is yet an open issue.
Pruet et al. (2004) constructed a simple model of hydrodynamic disk winds and
argued that in such winds the electron fraction typically evolves to $Y_{e}\simgt0.5$.  They also
pointed out that lower values of the asymptotic electron fraction are expected
in centrifugally driven winds.   Their model assumes effectively that the outflow is ejected from
a hydrostatic corona, where the density is high and the flow velocity is very small.
As proposed by Beloborodov (2003), turbulent mixing can lift up neutron-rich matter from the inner disk
layers to its surface on a timescale much shorter than the neutronization time.   A fraction of this circulating material
may then be ejected as an outflow, in which case the conditions at the base of the flow may be vastly different
than those expected in the case of an outflow from a hydrostatic corona.

In a recent paper (Levinson 2006; hereafter LE06) a model for GRMHD outflow from a neutrino-cooled disk has been developed
and employed to calculate the structure of the outflow in the sub-slow magnetosonic
region and the mass loading of the outflow.   The method used to compute the mass flux is outlined in
\S  \ref{sec:model} below.  The principle conclusion drawn in that paper is that ejection of relativistic outflows from the
innermost disk radii, within several $r_g$ or so, is possible in principle for certain magnetic field
configurations even in non-rotating black holes, provided the neutrino luminosity is
sufficiently low, $L_\nu\simlt10^{52}$ erg s$^{-1}$, and the magnetic field is sufficiently strong, $B\sim10^{15}$ G.
The conditions found to be optimal for the launching of an ultra-relativistic jet are also the conditions
favorable for large neutron-to-proton ratio in the disk.  However, the composition profile of the outflow
was not computed in LE06, and the question whether the outflow can retain a large degree of neutronization was left open.
In this paper we incorporate the evolution equation for the electron fraction into the model developed in
LE06, and compute numerically the structure and nuclear composition of the outflow beneath the slow
magnetosonic point for a range of conditions in the disk.  We specifically consider outflows emanating from a
hydrostatic disk corona and steady outflows driven by disk turbulence.

\section{GRMHD Disk Outflow Model}
\label{sec:model}

We consider a stationary, axisymmetric MHD wind expelled from the
surface of a hot, magnetized disk accreting onto a Schwarzschild
black hole.  The range of conditions in the disk is envisaged to be
similar to that computed by  CB07 for a black hole of mass
$3M_\odot$, accretion rates $\dot{M}_{\rm acc}\sim10^{-2} - 1$
$M_{\odot}$ s$^{-1}$ and viscosity parameters $\alpha_{\rm vis}=0.01
- 0.1$. Those authors generalized earlier work by Popham et. al.
(1999) and Pruet et al. (2003) to incorporate calculations of
electron degeneracy and nuclear composition self-consistently, which
affect the conditions in and neutrino emission from the disk. Under
the conditions envisioned above the dominant cooling mechanism in
the disk is neutrino emission, with a total luminosity of
$L_\nu+L_{\bar{\nu}}\simeq 0.04\dot{M}_{\rm acc}c^2$ for accretion
rates above the ignition values (CB07). The neutrino luminosity is
considerably higher if the black hole is rapidly rotating. The major
fraction of the neutrino luminosity is generated in the inner disk
regions, within 10 Schwarzschild radii or so, although for very low
$\alpha_{\rm viss}$ the neutrino emission may extend to much larger
radii. The matter in the inner disk regions is typically neutron
rich.  The electron fraction $Y_e$ decreases, quite generally, with
increasing $\dot{M}_{\rm acc}$ and decreasing $\alpha_{\rm vis}$.
For example, for $\dot{M}_{\rm acc}=0.2$ $M_{\odot}$ s$^{-1}$ the
electron fraction changes from $Y_e\simeq 0.03$ to $Y_e\simeq 0.15$
as $\alpha_{\rm vis}$ is varied from 0.01 and 0.1 (CB07).

\subsection{Basic Flow Equations}
The model outlined in LE06 calculates the structure of the GRMHD outflow below the slow magnetosonic point
for a given magnetic field geometry,
treating the neutrinos emitted from the disk as an external energy and momentum source.
To simplify the analysis the neutrino source has been taken to be spherical with radius $R_\nu$.
The model is characterized by three parameters: the black hole mass $M_{BH}$,
the neutrino flux $L_{\nu}/R_\nu^2$, and the mean neutrino energy $<E_{\nu}>$.
It solves a set of coupled ODEs that are derived from the general
relativistic  energy-momentum equations, describing the change along a
given streamline, $\Psi(r,\theta)=$ const,  of the specific energy ${\it E}$ ,
specific entropy $s$, and poloidal flow velocity $u_p$:
\begin{eqnarray}
(\rho/m_N)k_BT s^\prime=-u_\alpha q^\alpha,\label{s}\\
\rho c^2 {\it E}^\prime=-q_t,\label{E=q}\\
(\ln u_p)^\prime={F\over D}\label{u_p}.
\end{eqnarray}
Here  $^\prime$ denotes the derivative along streamlines,
$u^\alpha\partial_\alpha$, $q^{\beta}$ denotes the source terms
associated with energy and momentum exchange with the external
neutrino source, $u^\alpha$ is the outflow 4-velocity, $\rho$ is the
baryon rest mas  density and $T$ is the temperature.   The
denominator on the right hand side of eq. (\ref{u_p}) is given explicitly as
$D=-(\alpha^2-R^2\Omega^2-M^2)^2(u_p^2-u_{SM}^2)
(u_p^2-u_{FM}^2)/u_A^2$, where $u_A$, $u_{SM}$ and $u_{FM}$ are the
Alfv\'en, slow and fast magnetosonic wave speeds, respectively,
$\alpha$ is the lapse function, $\Omega$ is the angular velocity
defined below, $R$ is the cylindrical radius and $M=u_p/u_A$ is the
Alfv\'en Mach number. The term $F$ can be expressed as
$F=\zeta_1(\ln B_p)^\prime+\zeta_2(\ln \alpha)^\prime+\zeta_3(\ln
R)^\prime+ \zeta_4(\ln {\it E})^\prime+\zeta_5 (\ln s)^\prime$,
where the coefficients $\zeta_k$ are functions of the flow
parameters, given explicitly in LE06, and $B_p$ is the
poloidal field component.  Since the derivatives $(\ln B_p)^\prime$
and $(\ln R)^\prime$ depend on the magnetic field geometry which is
unknown a priori, additional equation is needed.  Our approach is to
invoke a given field geometry.  To examine the dependence of mass
flux on the latter, we obtained solutions for different magnetic
field configurations, focusing particularly on split monopole and
$r$ self-similar geometries.   Equations (\ref{s})-(\ref{u_p}) are
augmented by an equation of state for the mixed fluid of baryons,
photons and electron-positron pairs.  In addition there are three
integrals of motion of the MHD system: the mass-to-magnetic flux
ratio $\eta(\Psi)$, the angular velocity of magnetic field lines
$\Omega(\Psi)$, and the specific angular momentum ${\it L}(\Psi)$.
The two invariants $\Omega(\Psi)$ and ${\it L }(\Psi)$ are fixed by
a choice of boundary conditions.  The mass flux $\eta(\Psi)$ is an
eigenvalue of the problem, and is determined by the regularity
condition at the slow magnetosonic point.

\subsection{The Electron Fraction}
\label{sec:elect-frac}
For the range of conditions considered below the matter in the sub-slow magnetosonic region consists mainly of free
nucleons.  The total baryon density is then given by $\rho=m_pn_p+m_n n_n$, where $n_n$ and $n_p$ denote
the number density of neutrons and protons, respectively, and $m_n$, $m_p$ the corresponding masses.
The neutron-to-proton ratio is related to the electron fraction $Y_e$ through
\begin{equation}
Y_e(T,\rho)=\frac{n_p}{n_p+n_n},
\end{equation}
and is determined by a competition between the following reactions:
\begin{eqnarray}
e^-+p\rightleftharpoons n+\nu_e,\label{react-1}\\
e^+ + n\rightleftharpoons p+\bar{\nu}_e.\label{react-2}
\end{eqnarray}
Neutron decay is negligible due to the long life-time of the neutron
compared with the characteristic timescales involves. Lepton capture on
heavy nuclei can be ignored since as stated above (and will be confirmed below) in the regime
considered here the matter consists mainly of free nucleons.  The reactions $\nu_e+\bar{\nu}_e\leftrightarrow
e^++e^-$ are typically unimportant at the characteristic densities
and temperatures involved, nonetheless, they are
incorporated for completeness in the source terms $q^\beta$ that
appear on the right-hand side of eqs. (\ref{s}) and (\ref{E=q}) (see
LE06 for details).

The change of the electron fraction along a given streamline is determined by
\begin{equation}
Y_e^\prime=\lambda_{\nu n}+\lambda_{e^+ n}-(\lambda_{\nu n}
+\lambda_{e^+ n} + \lambda_{\bar{\nu} p} +\lambda_{e^-
p})Y_e,\label{Y_e-evolution}
\end{equation}
where $\lambda_{e^-p}$,  $\lambda_{e^+n}$,  $\lambda_{\bar{\nu}p}$,
$\lambda_{\nu n}$, are the rates for the forward and reverse
reactions in eqs. (\ref{react-1}) and (\ref{react-2}), and $'$
denotes again the derivative $u^\alpha\partial_\alpha$.

The rates for electron and positron capture are given by (Fuller et al. 1980),
\begin{eqnarray}
\lambda_{e^-p}=k \int_{\Delta /m_e}^\infty w^2(w-\Delta /m_e)^2G_-(1,w) S_-(1-S_\nu)dw, \\
\lambda_{e^+n}=k \int_{1}^\infty w^2(w+\Delta /m_e)^2G_+(1,w)S_+(1-S_{\bar{\nu}})dw
\end{eqnarray}
where $k\simeq 6.414\times 10^{-4}$ s$^{-1}$  is comparative
half-life related to the Gamow-Teller and Fermi matrix elements,
$\Delta =m_n-m_p=1.293$ MeV is the neutron-proton mass difference,
$m_e$ is the electron mass, $w$ is the total energy (rest mass and
kinetic energy) in units of $m_ec^2$, $S_-$ and $S_+$ are the
electron and positron distribution functions, respectively, and
$G_\pm(Z=1,w)$ are the Coulomb correction factors, given in Fuller et al. (1980).
For the range of temperatures and densities considered the e$^\pm$
pairs and photons are in perfect thermodynamic equilibrium in the
sub-slow magnetosonic region, owing to the huge optical depth there.
The functions $S_\pm$ are then given by $S_\pm(T,\mu)=[\exp\{(w\pm
\mu)/\theta\}+1]^{-1}$, where $\theta=k_BT/m_ec^2$ and $\mu$ is the electron chemical potential
(measured in units of $m_ec^2$).  The latter is determined from the charge
neutrality condition:
\begin{equation}
\frac{\rho}{m_p}Y_e=n_{-} -
n_{+}=\frac{1}{\pi^2}\left(\frac{m_ec}{\hbar}\right)^3\int_0^\infty{[S_-(T,\mu)-S_+(T,\mu)]p^2dp},
\end{equation}
with $p=(w^2-1)^{1/2}$ being the electron/positron momentum in units of $m_ec$.

As mentioned above, we assume for simplicity that the neutrino source is spherical with some characteristic luminosity $L_\nu$, mean neutrino energy $<E_{\nu}>$, and radius $R_\nu$.  In cases where the disk is $\nu$-opaque $R_\nu$ denotes the radius of the corresponding $\nu$-sphere.  Due to the potentially different opacities for $\nu_e$ and $\bar{\nu}_e$ absorption the luminosity $L_{\bar{\nu}}$, mean energy $<E_{\bar{\nu}}>$ and photospheric radius $R_{\bar{\nu}}$ of the antineutrinos may differ from that of neutrinos.  The details depend on the temperature profile in the neutrino production zone, which is governed by the dissipation mechanism in the disk.  In $\nu$-transparent disks we naively anticipate $L_\nu/L_{\bar{\nu}}=<E_{\nu}>/<E_{\bar{\nu}}>=R_\nu/R_{\bar{\nu}}=1$.  The rates for the reactions $\nu_e+n\rightarrow p+e^-$ and
$\bar{\nu}_e+p\rightarrow n+e^+$ at the base of the outflow can be approximated as (Qian and Woosley 1996)

\begin{eqnarray}
\lambda_{\nu n}\simeq \frac{1+3\alpha^2}{2\pi^2}G_F^2\frac{L_\nu}{R_\nu^2}
\left(\epsilon_\nu+2\Delta+\frac{\Delta^2}{<E_\nu>}\right)(1-x_\nu), \label{lambda1}\\
\lambda_{\bar{\nu} p}\simeq \frac{1+3\alpha^2}{2\pi^2}G_F^2\frac{L_{\bar{\nu}}}{R_{\bar{\nu}}^2}
\left(\epsilon_{\bar{\nu}}-2\Delta+\frac{\Delta^2}{<E_{\bar{\nu}}>}\right)(1-x_{\bar{\nu}}),\label{lambda2}
\end{eqnarray}
where $\alpha\simeq1.26$, $G_F^2=5.29\times 10^{-44}$ cm$^2$ MeV$^{-2}$ is
the Fermi coupling constant, $x_{\nu(\bar{\nu})}=(1-R_{\nu(\bar{\nu})}^2/r^2)^{1/2}$ is a
geometrical factor, and $\epsilon_{\nu(\bar{\nu})}=
<E_{\nu(\bar{\nu})}^2> / <E_{\nu(\bar{\nu})}>$. Detailed
calculations of the neutrino spectrum emitted by the disk is beyond
the scope of this paper. In what follows we shall assume for
simplicity that the $\nu_e$ and $\bar{\nu}_e$ are emitted from the same
region with the same luminosity and spectrum, that is, we take
$R_\nu=R_{\bar{\nu}}$,  $L_\nu=L_{\bar{\nu}}$ and  $<E_\nu>=<E_{\bar{\nu}}>$.
The energy moments $<E_{\nu}>$ and $\epsilon_{\nu}$ depend
on the shape of the neutrino spectrum. For a blackbody spectrum
$<E_{\nu}>=3.15 k_B T_\nu$, $\epsilon_{\nu}=4.1 k_B T_\nu$, where
$T_\nu$ denotes temperature of the neutrino source, whereas for a
neutrino transparent source $<E_{\nu}>\simeq 5 k_B T_\nu$,
$\epsilon_{\nu}\simeq 6 k_B T_\nu$ (Beloborodov, 2003).  The free
parameters $L_\nu$, $T_\nu$, and $R_\nu$ are estimated by employing
the results of CB07.

\subsection{Conditions at the Flow Injection Point}
In LE06 it has been assumed that the outflow connects to a hydrostatic disk corona where the density
is high and the entropy per baryon,
\begin{equation}
s=8.7+\ln(T_{MeV}^{3/2}/\rho_{9})+0.53(T_{MeV}^3/\rho_{9}),\label{s-initial}
\end{equation}
is relatively small.  Under this assumption, the integration of eqs. (\ref{s})-(\ref{u_p})
starts at sufficiently dense disk layers where the light fluid pressure, $p_l=(11/12)aT^4$, roughly equals the
baryonic pressure, $p_b=\rho k_BT/m_p$.  The mass flux $\eta$
is then adjusted iteratively by changing the boundary value of the poloidal velocity $u_{p0}$, until a smooth transition
across the slow magnetosonic point is obtained.

Different boundary conditions may apply if the disk is turbulent.  Beloborodov (2003) and  Pruet et al. (2003) proposed that turbulent mixing may quickly lift up neutron-rich matter from the inner disk layers to its surface, and that a small portion of circulating material will then be picked up by the outflow.  Since vertical mixing is expected to occur over the sound crossing time, $t_{\rm mix}\simeq H/c_s\simeq \Omega_K^{-1}$ where $H$ is the disk scale height, which for disk temperatures $T_d<8$ MeV is smaller than both, the neutronization time and neutrino cooling time (Beloborodov 2003), we anticipate the temperature and electron fraction at the outflow injection point to be roughly equal to their values at the disk midplane.  The disk turbulence is expected to be subsonic and, therefore, the poloidal velocity $u_{p0}$ of circulating matter which is ejected into the flow is anticipated to be a fraction of the sound speed inside the disk, viz., $u_{p0}\simlt c_s\simeq0.03 c T_{\rm MeV}$.  This is typically below the slow magnetosonic speed at the flow critical point (as will be confirmed below).  Consequently, under the assumption that the outflow above the turbulent layer is steady (averaged over times longer than the mixing time) it should
pass through the slow magnetosonic point.  This again determines the mass flux, as in the case of outflows that connect to a hydrostatic corona.  However,
in this scenario the outflow starts from a layer of much smaller density and much higher specific entropy than in the solutions obtained in LE06.  To account for such cases, we relax the
constraint imposed in LE06 on the specific entropy at the flow injection point.  Specifically, for any given choice of initial temperature, $T_0=T(r=r_0)$, and electron fraction, $Y_{\rm e0}=Y_{\rm e}(r=r_0)$, we construct a family of transonic solutions that are characterized by one parameter: the initial entropy $s_0=s(T_0,\rho_0)$ (or equivalently the initial density $\rho_0$).

\section{Analytic Treatment}
\label{sec:Analy}
In regions where the outflow is radiation dominated, viz., $p_{l}/p_b>1$,  the density satisfies
\begin{equation}
\rho_9<0.13 T_{MeV}^3.
\end{equation}
The degeneracy condition $\rho_9>0.2 T_{MeV}^3$ implies that in this
region of the flow the electrons are non-degenerate.   Under such
conditions the capture rates for electrons and positrons are given
approximately by $\lambda_{e^-p}\simeq \lambda_{e^+n}\simeq0.45
T_{MeV}^5$ s$^{-1}$ provided $T_{MeV}\simgt1$ (Qian \& Woosley
1996).  Using eqs. (\ref{lambda1}) and   (\ref{lambda2}) we then
obtain
\begin{equation}
\frac{\lambda_{\bar{\nu} p}}{\lambda_{e^- p}}\simeq
\frac{\lambda_{\nu n}}{\lambda_{e^+ n}}\simeq
2\times 10^2\;L_{\nu52}\;\epsilon_{\nu,MeV}\;R^{-2}_{\nu6}\;T_{MeV}^{-5}\;(1-x_\nu).
\end{equation}
The latter ratio can also be expressed in terms of the neutrino heating rate,
$q=5L_{\nu52}\epsilon^2_{\nu,MeV}R_{\nu6}^{-1}(1-x_\nu)$ MeV s$^{-1}$ baryon$^{-1}$, as
\begin{equation}
\frac{\lambda_{\nu n}}{\lambda_{e^+ n}}\simeq
40\;\left(\frac{q}{\rm 1\;MeV\;s^{-1}\;baryon^{-1}}\right)\;\epsilon^{-1}_{\nu,MeV}\;T_{MeV}^{-5}.
\end{equation}
For accretion rates above the ignition value
$L_\nu=(L_\nu+L_{\bar{\nu}})/2\simeq0.02 \dot{M}_{\rm acc}c^2$
in case of accretion onto a $3M_\odot$ Schwarzchild black hole (CB07).
Adopting for illustration $\dot{M}_{\rm acc}=0.2 M_\odot$
s$^{-1}$ and $R_{\nu6}=5$ yields $L_{\nu52}/R^2_{\nu6}\simeq0.03$. For
the latter choice of $\dot{M}_{\rm acc}$ CB07 obtained a peak
temperature of $T_\nu=2.5$ MeV for $\alpha_{\rm viss}\simlt0.03$ and
$T_\nu=3$ MeV for $\alpha_{\rm viss}=0.1$.  With $\epsilon_{\nu,MeV}=5T_{\nu,MeV}$ we
then have at the base of the flow $\lambda_{\nu n}/\lambda_{e^+ n}=0.75$ and 0.35,
for $\alpha_{\rm viss}\simlt0.03$ and $\alpha_{\rm
viss}=0.1$, respectively.  However, as the flow decelerates the
temperature drops and this ratio quickly rises.  For the range of
disk parameters considered here, $L_{\nu52}\simgt0.2$,
$\epsilon_{\nu,MeV}>10$, $T_{0,MeV}\simlt3$, and $R_{\nu6}\simeq3$,
we find, quite generally, that the asymptotic electron fraction is
governed by neutrino capture.

Denoting by $dl$ the proper length along a streamline in
units of $R_\nu$, and neglecting $e^\pm$ capture, we can write eq.
(\ref{Y_e-evolution}) as
\begin{equation}
\frac{dY_e}{dl}=\frac{\lambda_{\nu n}R_\nu}{u_p}(1-Y_e)-\frac{\lambda_{\bar{\nu} p}R_\nu}{u_p}Y_e.
\label{dY/dl-b}
\end{equation}
Adopting $\epsilon_{\nu}=5 k_BT_\nu$ and using eq. (\ref{lambda1}) we obtain at the flow injection point,
\begin{equation}
\frac{\lambda_{\nu n}R_\nu}{u_{p}}\simeq
0.02c\;L_{\nu52}\;T_{\nu,MeV}\;R_{\nu6}^{-1}\;u_{p0}^{-1},\label{lambdat_f}
\end{equation}
where c is the speed of light.  From eq. (\ref{dY/dl-b}) it is evident that the asymptotic value of
the electron fraction will remain near its initial value provided the outflow time, $t_f\sim R_\nu/u_{p0}$,
is shorter than the neutronization time, $\lambda_{\nu n}^{-1}$, which requires
\begin{equation}
u_{p0}> 0.02c L_{\nu52} T_{\nu, MeV}R_{\nu6}^{-1}.\label{u_p_condition}
\end{equation}
In the following, it is shown that this condition is satisfied in the case of
centrifugally driven winds and outflows driven by disk turbulence.  In thermally driven winds
the outflow time is much longer than that implied by condition (\ref{u_p_condition}).  The electron fraction then
quickly evolves to its equilibrium value, $Y_{\rm e,eq}\simeq \lambda_{\nu n}/(\lambda_{\nu n}+\lambda_{\bar{\nu} p})$.  For
a symmetric $\nu_e$ and $\bar{\nu}_e$ emission eqs. (\ref{lambda1}) and (\ref{lambda2}) yield
$\lambda_{\nu n}>\lambda_{\bar{\nu} p}$ due to the threshold effect,
resulting in asymptotic proton excess ($Y_{\rm e,\infty}>0.5$) in those outflows.  Our detailed calculations yield $Y_{\rm e,\infty}\simgt0.5$ also
in cases where $e^\pm$ capture dominates the evolution of the electron fraction in the wind.

At the slow magnetosonic point the sound speed $a_{sc}$ is equal to a modified escape speed.
Typically $a_{sc}\sim0.1 c$ (LE06).  Since $a_s^2\simeq 4p_l/3\rho c^2$ in the subslow region it implies
that the density at the injection point must satisfy $\rho_0\simgt10^7 T_{MeV}^4$ gr cm$^{-3}$.  This, in turn, implies a
(one sided) mass flux of
\begin{equation}
\dot{M}\simeq \pi r_0^2 \rho u_{p0}\simgt10^{30} T^4_{MeV}(u_{p0}/0.1\;c)\qquad {\rm gr\ cm^{-3}}
\label{mdot-analy}
\end{equation}
for an injection radius $r_0\simeq R_\nu=3\times10^6$ cm.
Those rough estimates are confirmed by numerical integration of the full set of equations.  The above considerations suggest that for
explosion energies inferred in GRBs neutron rich outflows can only have modest Lorentz factors ($\Gamma\sim$ a few), unless somehow
the turbulent mixing process manage to selectively accelerate only a very small fraction of the neutron rich matter that is lifted to the disk surface to velocities well in excess of the escape velocity.

\section{Numerical Results}
Equations (\ref{s})-(\ref{u_p}), (\ref{Y_e-evolution})-(\ref{lambda2}) have been
integrated numerically, as in LE06, using a split monopole magnetic field geometry.
Each field line in this configuration is characterized by two parameters: the inclination angle $\theta$ of the field
line, here measured with respect to the symmetry axis, and the radius $R_0$ at
which the field line intersects the equatorial plane.   In the following we consider, in turn, outflows from a steady disk and
outflows driven by disk turbulence.

\subsection{Steady Outflow from a Hydrostatic Disk Corona}
\label{sec:hydro-corona}
The integration of these models starts in the dense disk layers, where the pressure is dominated by the baryons.  The
initial temperature $T_0$ is taken to be equal to the temperature of the neutrino source, viz., $T_0=T_\nu$.
For a given choice of $T_\nu$ the density at the origin, $\rho_0$, is chosen such that
the light fluid pressure, $p_l=(11/12)aT_0^4$, does not exceed the baryonic pressure, $p_b=\rho_0 k_BT_0/m_p$.
The initial entropy $s_0=s(T_0,\rho_0)$ is then determined using eq. (\ref{s-initial}).
To verify that the solution is insensitive to our choice of boundary conditions,
each integration has been repeated several times for a given choice
of our model parameters, each time with a different value of $\rho_0$.
We find that as long as $p_l(T_0)<p_b(T_0,\rho_0)$ the results are indeed highly insensitive to our choice of $\rho_0$,
except for the initial value of the electron fraction which at these densities equals its equilibrium value.
For our fiducial model we choose the electron fraction at the injection point to be $Y_{e0}=0.1$, which is
the equilibrium value at a density $\rho\simeq10^{10}$ gr cm$^{-3}$ and temperature $T\simeq 2$ MeV (Beloborodov, 2003).

In general there are two distinct regimes.  The regime of unstable equilibrium corresponds to
magnetic field lines having inclination angles $\theta>\pi/6$.
Along such field lines the outflow is centrifugally driven and can be initiated even
in the cold fluid limit (Blandford \& Payne 1982; see LE06 for a generalization of this result to
the general relativistic case).  The slow magnetosonic point in this case is located very close
to the disk surface, at $z_{sm}<<r_0$, where the density is high.  Because of the high critical density the mass flux along such field lines is large and depends only weakly on the neutrino luminosity emitted from the disk.  For reasonable magnetic field strengths those outflows are typically subrelativistic.  The regime of stable equilibrium  corresponds to field line inclination angles $\theta<\pi/6$.
Along such field lines the mass flux is thermally driven, similar to the case of a spherical wind, and is a sensitive function of the neutrino luminosity.  The slow magnetosonic point in this regime is located higher above the disk, at heights $z_{sm}\sim r_0$ (see fig. 1).  The heating of the wind by the escaping neutrinos results in a steep rise
of the entropy per baryon, $s$, during the initial acceleration phase, after which it
saturates.  The asymptotic value of $s$ is larger for lower $L_\nu/R^2_\nu$, but does not seem to reach
extreme values ($s\simlt100$).  Adiabatic cooling is important along field lines of sufficiently small $\theta$,
rendering the critical density and temperature substantially smaller than in the case of centrifugally driven winds.
As a result the mass flux is strongly suppressed.   For $\theta<15^\circ$ the outflow is magnetically dominated
if the strength of the poloidal magnetic field at the disk surface is $B_p\sim10^{15}$ Gauss, as often envisioned.
The magnetic energy per baryon depends sensitively on the inclination angle of the field line near the surface
and can reach very large values (LE06).

Examples of thermally driven winds are exhibited in fig. 1, where the flow parameters are plotted
as functions of the height above the disk midplane, for $\tan \theta=0.2$ and $r_0=R_\nu=3r_g$.
The left panel corresponds to a disk temperature $T_0=2$ MeV and neutrino luminosity
$L_\nu=6\times 10^{51}$ erg s$^{-1}$, and the right panel to a disk temperature $T_0=3$ MeV and neutrino luminosity
$L_\nu=2\times 10^{52}$ erg s$^{-1}$.  The initial density and electron fraction in both examples are $\rho_0=10^{10}$
gr cm$^{-3}$ and  $Y_{e0}=0.1$, respectively.   The disk is transparent in both cases, and so the average neutrino energy is taken to
be $<E_\nu>\simeq 5kT_0$.  The mass flux, as defined in eq. (\ref{mdot-analy}), is found to be about $10^{28}$ gr s$^{-1}$ in
the former case and $10^{30}$ gr s$^{-1}$ in the latter, corresponding to a net energy per baryon of ${\it E}=500$ and  ${\it E}=5$, respectively, for a poloidal magnetic field strength of $B_{p0}=10^{15}$ Gauss.  As stated above, the solution is highly insensitive to the
choice of initial density $\rho_0$, provided $p_l/p_b<1$.  Changing the latter from $10^{10}$ to $10^9$ gr cm$^{-3}$ altered the mass
flux and asymptotic $Y_e$ by less than one percent.  It also had little effect on the profiles of the flow quantities.

From fig. 1 it is seen that the electron fraction reaches its final equilibrium value close to the base of the flow.  At the origin, the rates for neutrino capture and e$^\pm$ capture are comparable and $Y_e$ adjusts instantaneously to its local equilibrium value, $Y_{\rm e,eq}(\rho_0,T_0)\simeq 0.3$.  As the density and temperature drop the ratio $\lambda_{\nu n}/\lambda_{e^- p}$ increases, leading to a further increase of $Y_e$ until it reaches its final equilibrium value $Y_{\rm e,\infty}\simeq \lambda_{\nu n}/(\lambda_{\nu n}+\lambda_{\bar{\nu}})\simgt0.5$ .
This rapid evolution of $Y_e$ is characteristic to all thermally driven outflows, and is a consequence of the fact that the flow time, $t_{f}=(r_0/c)\int{dl/u_p}\simeq R_\nu/u_{p0}$, implied by the critical condition is considerably
longer than the neutronization timescale $\lambda_{\nu n}^{-1}$.
In the examples depicted in fig. 1 the poloidal velocity at the flow injection point is $u_{p0}=4.5\times10^{-7}$ (left panel)
and $u_{p0}=4.5\times10^{-5}$ (right panel), well below the
value required to retain low $Y_e$ (see eq. [\ref{u_p_condition}]).
The asymptotic electron fraction depends to some extent on the disk temperature $T_\nu$, owing to the threshold effect
(i.e., the dependence of the rates $\lambda_{\nu n}$ and $\lambda_{\bar{\nu} p}$ in eqs. [\ref{lambda1}] and [\ref{lambda2}] on
the neutron-proton mass difference $\Delta$).  In general, $Y_{\rm e, eq}$ tends to approach 0.5 as $T_\nu$ increases.  It should be kept in mind, however, that the uncertainty in the luminosity and mean energy ratios
of $\nu$ and ${\bar{\nu}}$ should be reflected in the final value of $Y_e$.

As stated in \S \ref{sec:elect-frac} above, our analysis ignores the presence of heavy nuclei.
As a consistency check, we plotted in fig. 1 the equilibrium value of the free nucleon mass
fraction, $X_{\rm free}(T,\rho)$, taken from Woosley \& Baron (1992) ($X_{\rm free}=1$ implies free nucleons only).
This estimate of $X_{\rm free}$ is accurate at the base of the flow, where the matter is in rough
kinetic equilibrium.  Further up, where the flow velocity exceeds values at which the expansion time becomes
shorter than the weak interaction timescale it underestimates the fraction of free nucleons.
As seen, our neglect of lepton capture on heavy nuclei in the relevant phase of the flow, where the change
in the electron fraction is significant, is justified.

Example of a centrifugally driven outflow is shown in fig. 2, for
two different inclination angles. As seen, the electron fraction
evolves rather slowly. This is a consequence of the short expansion
time in the sub-slow region.  In general, the asymptotic electron
fraction would depend on the structure of the flow above the slow
magnetosonic point.  However, given the high poloidal velocity at
the critical point ($u_{pc}\sim0.04 c$ in both cases shown in fig.
1) we expect only little evolution above the slow point. The general
conclusion is that centrifugally driven outflows can retain low
values of $Y_e$.   This trend is consistent with that found by Pruet
et al. (2004). The angular velocity of magnetic field lines in those
calculations is $\Omega= 0.95\Omega_K$, where
$\Omega_K=(r_g/2r_0^3)^{1/2}$ is the Keplerian angular velocity at
the disk midplane. The corresponding mass fluxes are
$\dot{M}\simeq0.1$ and $0.2\; M_\odot$ s$^{-1}$, for
$\theta=30^\circ$ and $\theta=60^\circ$, respectively.  Such high
mass loss rates imply that a significant fraction of the accreted
matter will be expelled from the disk before reaching the black
hole, which should affect the disk structure considerably.  In
particular, the disk is likely to become sub-Keplerian, and this
would feed back on the flow.  As illustrated in LE06, the mass flux
is highly sensitive to the choice of $\Omega$. For example,
repeating the calculations in fig.1 with $\Omega=0.85\;\Omega_K$
yielded a mass flux smaller by about an order of magnitude.
Interestingly, the location of the slow magnetosonic point was not
altered at all, and the poloidal velocity at the critical point
changed by only $50\%$.  The reason is that the outflow accelerates
faster.  As a consequence the asymptotic electron fraction has not
changed significantly.

\subsection{Steady Outflow from a Turbulent Disk}
As explained above, this model assumes that turbulent mixing lifts neutron rich matter from the inner disk layers
to the surface on a timescale much shorter than the neutronization time.
The temperature $T_0$ and electron fraction $Y_{e0}$ at the flow injection point are then taken to be equal
to their values at the disk midplane, as before.  However, the entropy per baryon at the injection point is
allowed in this model to be much higher than in the case of outflows that emerge from a hydrostatic corona,
and is treated essentially as a free parameter.  The only restriction imposed on the solution is that it should
start sub-slow, that is, the slow magnetosonic Mach number at the flow injection
point must be be smaller than unity.

In a turbulent disk the magnetic field is likely to be disordered.   In that case the stream function should be considered
as defining the direction of streamlines.   We stress that our calculations are applicable
also to hydrodynamic flows.   It is worth noting that a flow can, in principle, be Poynting flux dominated
even in case of disordered magnetic fields (e.g., Proga et al. 2003), although, as discussed below, for the outflows
considered here the magnetic energy per baryon in typically smaller than unity.

Sample results are shown in fig. 3.  As expected, the evolution of the electron fraction is suppressed as
the initial Mach number of the injected flow is increased.
We find neutron excess ($Y_{e\infty}<0.5$) in solutions for which
the slow magnetosonic Mach number at the flow injection point, $M_{\rm SM}$, exceeds 0.1 roughly.
At the same time the mass flux also increases.  We find that
solutions that retain $Y_e<0.5$ have a relatively large mass flux, $\dot{M}\simgt10^{30}$ gr s$^{-1}$, in accord with
eq. (\ref{mdot-analy}).  We therefore conclude that neutron rich outflows expelled from a turbulent disk
are likely to be sub or at best mildly relativistic.  Note that neutrino heating is unimportant (negligible change in
specific entropy in all the solutions exhibited in fig. 3).  This is because of
the relatively short expansion time at the base of the flow, $t_f\lambda_{\nu n}<1$.
Thus, the structure of those outflows depends on the disk temperature,
but is independent essentially of the neutrino luminosity ( $Y_e$ does depend on $L_\nu$ of course).

\section{Summary and Conclusions}
We have computed the structure and the neutron-to-proton ratio in GRMHD outflows from
a neutrino-cooled disk accreting onto a Schwarzschild black hole.   We considered both, outflows emanating from
a hydrostatic disk corona and steady outflows driven by disk turbulence.  The main results are:

1. In thermally driven outflows that emerge from a steady disk the
neutronization time at the base of the outflow is much shorter then
the outflow time.  In this case the electron fraction quickly
evolves to its equilibrium value which, for the parameter regime
explored above, is dominated by neutrino capture.  For
$\nu$-transparent disks this implies asymptotic proton excess
($Y_e>0.5$).  The mass flux driven by neutrino and viscous heating
along magnetic field lines inclined at small angles to the vertical
($\theta\simlt15^\circ$) is found to be rather small for low
viscosity disks ($\alpha_{\rm viss}\simlt0.03$) and moderate
accretion rates ($\dot{M}_{\rm acc}\simlt0.1$).  If the outflow
energy is extracted magnetically from the disk with a luminosity
comparable to that observed in GRBs, then it can in principle
accelerate to a high Lorentz factor.  Luminous disks
($L_\nu>10^{52}$ erg s$^{-1}$, $T_\nu\simgt3$ MeV) give rise to a
much larger mass loss rate ($\dot{M}\simgt10^{-3}$ $M_\odot$
s$^{-1}$). Such outflows can in principle synthesize the amounts of
$^{56}$Ni required to explain the lightcurves of the associated SN
(Pruet et al. 2004).

2. Centrifugally driven winds can retain large neutron excess by virtue of the much shorter expansion time.  In those outflows the slow
magnetosonic point occurs very close to the disk surface, where the density is high.  The large critical density results in a substantial baryon loading of the flow, rendering it subrelativistic.  The large mass loss rates obtained imply that those outflows should
considerably affect the disk structure (this conclusion may hold true in general if magnetic extraction of angular momentum is significant). A complete treatment requires self-consistent solutions of the disk and the outflow.

3. Outflows driven by disk turbulence can also retain large neutron
excess if the slow magnetosonic Mach number at the origin satisfies
$M_{\rm SM}\simgt0.1$. In this scenario a fraction the circulating
material is ejected into the flow over a timescale much shorter than
the neutronization time (Beloborodov 2003; Pruet et al. 2003).  The
in-disk temperature and composition are then expected to be
preserved at the base of the flow.   As a result, the asymptotic
entropy per baryon is larger than in outflows from a steady
disk (see fig. 3). The poloidal velocity at the flow injection point
is likely to be a fraction of the sound speed inside the disk, which
is typically large enough to suppress the evolution of the electron
fraction in the outflow.  The mass flux carried by such outflows is
found to be relatively large, on the order of $10^{-3} M_\odot$
s$^{-1}$.

The main conclusion is that neutron rich outflows can be expelled from neutrino-cooled disks under certain conditions.  However, those
outflows are expected to be sub or at best mildly relativistic, at least in cases where the central black hole is non rotating.
Relativistic outflows can, in principle, be launched magnetically in the polar region, but those are, in general, proton rich.
The above results may be altered if the central black hole is rapidly rotating.

The slow, neutron rich winds discussed above may play an important role in the collimation of the central GRB-producing jet, and may also
provide a contaminating baryon source (Levinson \& Eichler 2003; McKinney 2006) for a central, baryon-free fireball that may be produced, e.g.,
via neutrino annihilation on magnetic field lines penetrating the horizon.   Baryon pick-up by the central jet may have some
interesting implications for fireball physics (e.g., Eichler \& Manis 2007).

This work was supported by an ISF grant for the Israeli Center for
High Energy Astrophysics

\newpage
\begin{figure}[h]
\centering
\includegraphics[width=15cm]{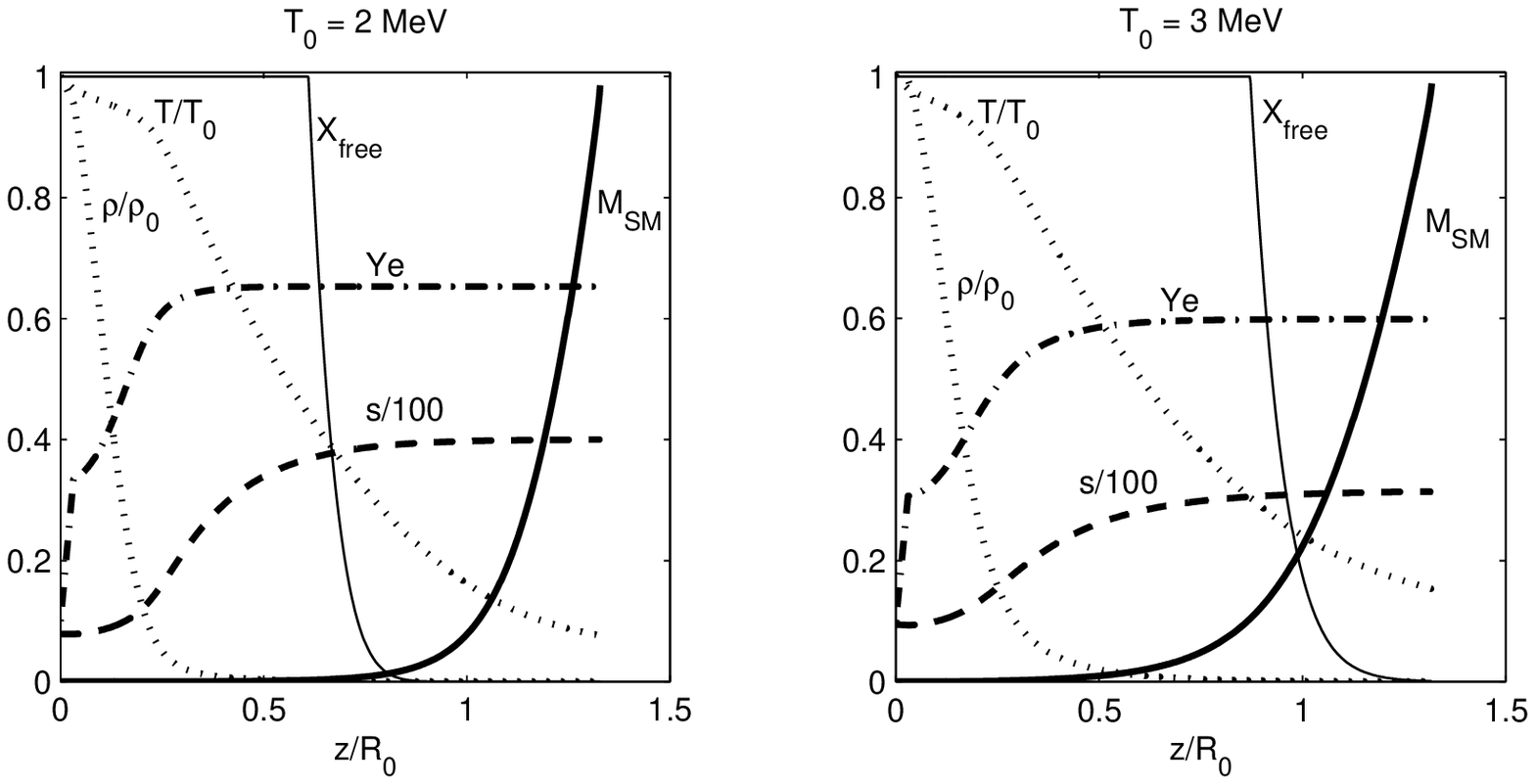}
\label{fig:T2T3_4_xfree} \caption {\small Flow parameters along a
streamline for a pressure driven flow, assuming $\tan\theta=0.2$,
surface magnetic field $B_{p0}=10^{15}$ G, $(B_\phi/B_p)_0=0.1$, and
initial density $\rho_0=10^{10}$ gr cm$^{-3}$. Each panel gives the
slow magnetosonic Mach number $M_{SM}$ (solid line), the
dimensionless entropy per baryon $s$ (dashed line), the normalized
temperature and density (dotted lines), the electron fraction $Y_e$
(dotted-dashed line), and the free nucleon mass fraction $X_{\rm
free}$, as functions of the normalized height above the disk
$z/R_0$. }
\end{figure}

\newpage
\begin{figure}[h]
\centering
\includegraphics[width=15cm]{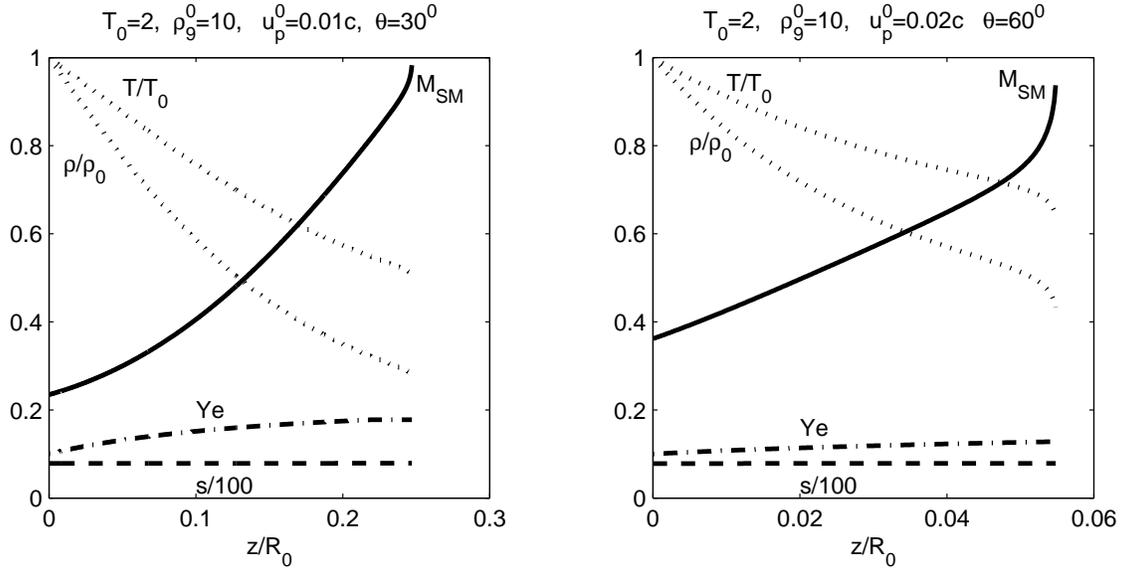}
\label{fig:CW}
\caption
{\small Same as figure 1 for a centrifugally driven outflow.}
\end{figure}

\newpage
\begin{figure}[h]
\centering
\includegraphics[width=15cm]{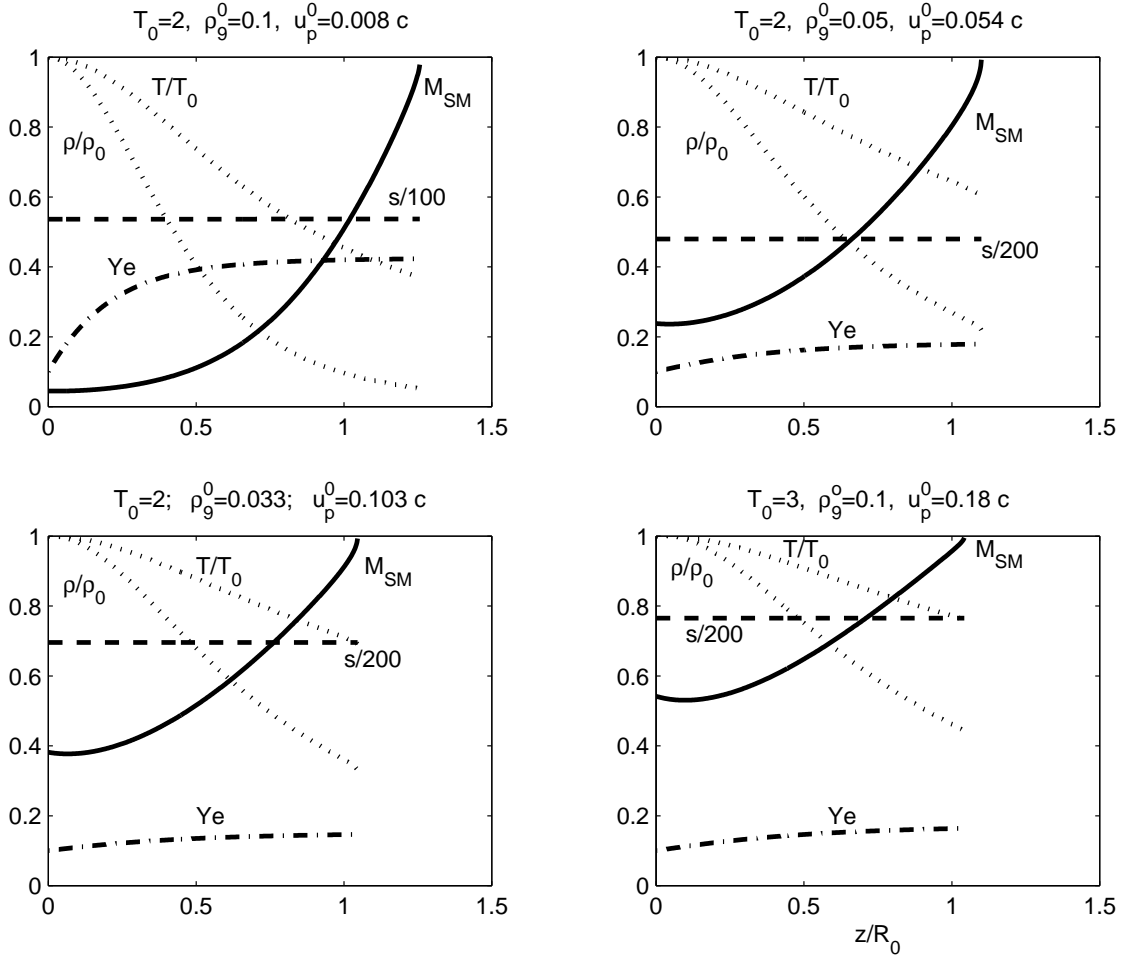}
\label{fig:T2T3_123}
\caption {\small Same as fig. 1 for outflow from a turbulent disk.
Initial values are indicated.  The corresponding mass fluxes, as
defined in eq. (\ref{mdot-analy}), are $\dot{M}=8\times10^{29}$,
$2.7\times10^{30}$, $3.5\times10^{30}$, $1.8\times10^{31}$ gr
s$^{-1}$ for the upper left, upper right, lower left and lower right
panels, respectively.}
\end{figure}

\end{document}